\documentstyle[12pt]{article}
\title{\huge  Wetting and layering transitions of a spin$-1/2$ Ising model in a random transverse field} 
\author{\bf  
 L. Bahmad, A. Benyoussef, A. El-Kenz and H. Ez-Zahraouy 
\\
 Laboratoire de Magn\'{e}tisme et de Physique
 des Hautes Energies
\\
Universit\'{e} Mohammed V, Facult\'{e} des Sciences \\
Avenue Ibn Batouta,  B.P. 1014, Rabat, Morocco
}
\date{ }
\begin{document}
\maketitle
  
\author{{\bf L. Bahmad, A. Benyoussef, A. El-Kenz and H. Ez-Zahraouy}
 Laboratoire de Magn\'{e}tisme et de la Physique des Hautes Energies \\
Universit\'{e} Mohammed V, Facult\'{e} des Sciences, Avenue Ibn Batouta,
B.P. 1014 \\
Rabat, Morocco}
\date{}
\maketitle

\begin{abstract}
\mbox{~~~  } The effect of a random transverse field (RTF) on the wetting and layering transitions of a spin-$1/2$ Ising model, in the presence of bulk and surface
fields, is studied within an effective field theory by using the differential operator technique. Indeed, the dependencies of the wetting temperature and wetting transverse field on the probability of the presence of a transverse field are established. For specific values of the surface field we show the existence of a critical probability $p_{c}$ above which wetting and layering transitions disappear.
\end{abstract}
 
\noindent ----------------------------------- \newline

\newpage

\section{Introduction}

\mbox{  }Recently much attention has been directed to study the wetting and layering transitions of magnetic surfaces Ising model. Experimental studies have motivated much theoretical works in order to understand and explain the growth of thin layers from only single atoms. A simple lattice-gas model with layering transitions has been introduced and studied in the mean field approximation by de Oliveira and Griffiths [1].
Multilayer films adsorbed on attractive substrates may exhibit a variety of possible phase transitions, as has been reviewed by Pandit {\it et al.} [2], Nightingale {\it et al.} [3], Patrykiejew {\it et al.} [4] and Ebner {\it et al.} [5-8]. One type of transitions is the layering transitions, in which the thickness of a solid film increases discontinuously by one layer as the pressure is increased. Such transitions have been observed in a variety of systems including for example $^{4}He$ [9,10] and ethylene [11,12] adsorbed on graphite. Ebner and Saam [13] carried out Monte Carlo simulations of such a lattice gas model. Huse [14] applied renormalization group technique to this model. It allowed the study of the effects on an atomic scale in the adsorbed layers. The lattice gas models applied to the wetting phenomena was reviewed by Dietrich [15]. The effect of finite size on such transitions has been studied, in a thin film confined between parallel planes or walls, by Nakanishi and Fisher [16] using mean field theory. \\
The model of transverse field was originally introduced by de Gennes [17] for hydrogen-bonded ferroelectrics such as $KH_{2}PO_{4}$. Since then, this model has been applied to several physical
systems, like $DyVO_{2}$, and studied by a variety of sophisticated
techniques [18-21]. 
The technique of differential operator
introduced by Kaneyoshi [22] is as simple as the mean field method and uses a
generalised but approximate Callen relation derived by Sa Barreto and
Fittipaldi [23]. The system has a finite transition temperature, which can be
decreased by increasing the transverse field to a critical value $\Omega _{c}
$. The effect of a transverse field on the critical behaviour and the
magnetisation curves was studied [18-21] and by Kaneyoshi {\it et al.} [24,25]. Using the perturbative theory, Harris {\it et al.} [26] have studied the layering transitions at $T=0$ in the presence of a transverse field.
Benyoussef and Ez-Zahraouy have studied the layering transitions of Ising model thin films using a real space renormalization group [27], and transfer matrix [28] methods. \\ 
On the other hand, the random systems have been known to be dominated by rare regions. This effect is particularly pronounced for random quantum systems at low or zero temperature far from critical points. Indeed, Griffiths [29] showed that the free energy is a non analytic function, because of rare regions. Having found all the derivatives being finite, Harris [30] concluded that this effect was very weak for classical systems.   
The simplest of all random quantum systems is the random transverse Ising model [31,32] (and references therein). \\
Using the mean field theory, we have studied in a previous work [33], the wetting and layering transitions of a spin$-1/2$ Ising model in the presence of a uniform transverse field.   
Our aim in this work is to study the effect of a random transverse field (RTF), on the wetting and layering transitions of a spin$-1/2$ Ising system, within an effective field theory (EFT) by using the differential operator technique.  
 The outline of this paper is as follows. In Section 2 we present the
formalism and the method. In Section 3 we investigate and discuss the phase diagrams.

\section{Model and method}

Transverse and longitudinal magnetic fields are applied to a system with N
coupled ferromagnetic layers. The Hamiltonian can be written as 
\begin{equation}
{\cal H}=-\sum_{<i,j>}J_{ij}S_{i}^{z}S_{j}^{z}-\sum_{i}(\Omega
S_{i}^{x}+H_{i}S_{i}^{z})
\end{equation}
where the first summation is carried out only over nearest-neighbour pairs of spins, $S_{i}^{\mu} ,(\mu=x,z)$ are the Pauli matrices of a spin$-1/2$ and $J_{ij}=J$ is the exchange interaction assumed to be constant. 
$H_{i}$ is the longitudinal field applied on the site $i$, assumed to be uniform in a layer $k$, and defined by: 
\begin{equation}
H_{k}=\left\{ 
\begin{array}{lll}
H+H_{s1} & \mbox{for} & k=1 \\ 
H & \mbox{for} & 1 < k < N \\ 
H+H_{s2} & \mbox{for} & k=N
\end{array}
\right.
\end{equation}
where the surface fields $H_{s1}$ and $H_{s2}=-H_{s1}$ are applied on the first layer $k=1$
and the last layer $k=N$, respectively. H is the longitudinal field. The
transverse field $\Omega$ is governed by the probability distribution law:
\begin{equation}
{\cal P}(\Omega)=p \delta(\Omega- \Omega_{0})+(1-p)\delta(\Omega)
\end{equation}
with $0 \le p \le 1$. The case $p=0$ corresponds to the absence of the transverse field, whereas $p=1$
is a situation with a uniform transverse field $\Omega = \Omega_{0}$. \\
We will use an effective field theory which neglects correlation between spins [35-36], but it takes into account the relations such as $<(s_{i})^{2}>=1$ exactly, where $<...>$ denotes the thermal average. This method has been used by several authors to study quantum systems [18,22-25], and disordered systems [18,19,21,37,38]. It is based on a cluster comprising a single selected site labelled $i$ and the neighbouring sites with which it directly interacts.
Hence, the part of the Hamiltonian containing the site $i$, is given by 
\begin{equation}
{\cal H}_{i}=(\sum_{j\ne i}J_{ij}S_{j}^{z}+H_{i})S_{i}^{z}+\Omega S_{i}^{x}
\end{equation}
the summation runs over nearest neighbour sites $j$ of the site $i$.
The diagonalization of the operator ${\cal H}_{i}$ leads to the eigen values 
$\lambda_{i}^{\pm}=\pm \sqrt{x_{i}^{2}+\Omega^{2}}$, with $x_{i}=\sum_{j\ne i}J_{ij}S_{j}^{z}+H_{i}$ and $J_{ij}=J$. \\
In the next we will use the notation $\sigma_{i}^{z}=2S_{i}^{z}$. Following Sa Barreto and Fittipaldi [23] we write the approximate relation:
\begin{equation}
<<\sigma_{i}^{z}>_{c}>=<\frac{Tr(\sigma_{i}^{z}\exp(-\beta 
{\cal H}_{i}))}{Tr(\exp(-\beta {\cal H}_{i}))}>
\end{equation}
where $\beta=1/(k_{B}T)$, $<...>_{c}$ indicates the mean value of $\sigma_{i}^{z}$ for a given
configuration $c$ of all other spins, $<...>$ denotes the average over all spin configurations. \\
Equations $(5)$ are not exact for an Ising system with a transverse field, nevertheless, they have been accepted as a reasonable starting point in many studies [22,23].
After averaging over the probability distribution $p$, the longitudinal magnetisation of a layer $k$, is given by 
\begin{equation}
m_{k}^{z}=p<f_{\Omega}(x_{k})>+(1-p)<f_{0}(x_{k})>
\end{equation}
with the functions $f_{\Omega}(x)$ and $f_{0}(x)$ are defined, respectively,
by 
\begin{equation}
f_{\Omega}(x)=\frac{x}{\sqrt{\Omega^2+x^2}}\tanh{\beta \sqrt{\Omega^2+x^2}}
\end{equation}
\begin{equation}
f_{0}(x)=\tanh{\beta x}
\end{equation}
Introducing the differential operator defined by: 
\begin{equation}
\exp{(\alpha D)}f_\mu(x)=f_\mu(x+ \alpha), (\mu = \Omega, 0)
\end{equation}
where $D=\frac{d}{dx}$, we can write the longitudinal magnetisation of a plane $k$ as 
\begin{equation}
m_{k}^{z}=p[\exp(<x_{k}>D)]f_{\Omega}(x)|_{x=0}]+(1-p)[\exp(<x_{k}>D)]f_{0}(x)|_{x=0}]
\end{equation}
\newline
For a cubic lattice model, each site is in interaction with 6 neighbours, one can write the magnetisation $m_{k}^{z}$ as 
\begin{equation}
\begin{array}{rll}
m_{k}^{z}= & p[(\cosh{JD}+<\sigma_{k}^{z}>\sinh{JD})^{4}(\cosh{JD}
+<\sigma_{k-1}^{z}>\sinh{JD})(\cosh{JD}+<\sigma_{k+1}^{z}>\sinh{JD}) &  \\ 
& (\exp(H_{k}D))f_{\Omega}(x)|_{x=0}]+(1-p)[(\cosh{JD}+<\sigma_{k}^{z}>\sinh{
JD})^{4} (\cosh{JD}+<\sigma_{k-1}^{z}>\sinh{JD}) &  \\ 
& (\cosh{JD}+<\sigma_{k+1}^{z}>\sinh{JD})(\exp(H_{k}D))f_{0}(x)|_{x=0}], & 
\end{array}
\end{equation}
with the boundary conditions $m_{N+1}^{z}=m_{0}^{z}=0$. \\
Neglecting the correlations, so that: 
\begin{equation}
<\sigma_{i}^{z}\sigma_{j}^{z}...\sigma_{l}^{z}>=<\sigma_{i}^{z}><
\sigma_{j}^{z}>...<\sigma_{l}^{z}>
\end{equation}
and using the differential operator, the final expression of the magnetisation $m_{k}^{z}$ is: 
\begin{equation}
\begin{array}{lll}
m_{k}^{z}= & B_{0}(p, \Omega)+B_{1}(p, \Omega)<\sigma_{k}^{z}>+B_{2}(p,
\Omega)<\sigma_{k}^{z}>^{2}+B_{3}(p, \Omega)<\sigma_{k}^{z}>^{3}
+B_{4}(p, \Omega)<\sigma_{k}^{z}>^{4} &  \\ 
& +<\sigma_{k-1}^{z}>(B_{5}(p, \Omega)+
B_{6}(p, \Omega)<\sigma_{k}^{z}>+B_{7}(p, \Omega)<\sigma_{k}^{z}>^{2}+B_{8}(p,
\Omega)<\sigma_{k}^{z}>^{3} &  \\ 
& +B_{9}(p, \Omega)<\sigma_{k}^{z}>^{4})+<\sigma_{k+1}^{z}>(B_{10}(p,
\Omega)+B_{11}(p, \Omega)<\sigma_{k}^{z}>+B_{12}(p, \Omega)<\sigma_{k}^{z}>^{2}
&  \\ 
& +B_{13}(p, \Omega)<\sigma_{k}^{z}>^{3}+B_{14}(p, \Omega)<
\sigma_{k}^{z}>^{4})+<\sigma_{k-1}^{z}><\sigma_{k+1}^{z}>(B_{15}(p,
\Omega)+B_{16}(p, \Omega)<\sigma_{k}^{z}> &  \\ 
& +B_{17}(p, \Omega)<\sigma_{k}^{z}>^{2}+B_{18}(p, \Omega)<
\sigma_{k}^{z}>^{3}+B_{19}(p, \Omega)<\sigma_{k}^{z}>^{4}) & 
\end{array}
\end{equation}
where $B_{l}(p, \Omega)=pA_{l}(\Omega)+(1-p)A_{l}(0), l=0,1,...,19$, and the
coefficients $A_{l}(\mu),(\mu = \Omega ,0), l=0,1,...,19;$ are given in the
Appendix A. \newline
Similarly, the transverse magnetisation $<\sigma_{k}^{x}>$, of a layer $k$, can be formulated
using a function $g_\mu(x)$ instead of the function $f_\mu(x)$, but in this
work, we are essentially interested in the wetting phenomena which is essentially 
governed by the longitudinal magnetisation $<\sigma_{k}^{z}>$.

\section{Results and discussion}

\mbox{~~~}In this section we present phase diagrams and longitudinal magnetisations of the model $(1)$ described in section 2. The ground state of this model was established in a previous work [33]. 
In the following we limit our calculations to $N = 20$ layers. The results are found to be similar for an arbitrary larger number of layers $N \geq 20$. We will distinguish between two cases, namely: \\
i) $p=1.0$: uniform transverse field, \\
ii) $0<p<1$: random transverse field. \\
The start point is a situation where all the spins are down. For fixed values of $H_{s1}/J$ and $\Omega/J$, there exist a temperature $T_{w}/J$ such that: \\
i) for $T/J<T_{w}/J$ the spins of all layers are down for $H/J<0$ and up once $H/J>0$, with the coexistence of the two cases at $H/J=0$, \\
ii) for $T/J>T_{w}/J$ and increasing the bulk field $H/J$
 the spins of the first layer will flip and become up: this is the surface  transition. Increasing the bulk field more and more, the spins of the second  layer will flip up, and so on. For a sufficiently large number of layers $N$, the complete wetting is reached when the number of layering transitions is close to the number of layers at $H/J=0$. $T_{w}/J$ is the wetting temperature. Similarly, the wetting transverse field $\Omega_{w}/J$ is obtained for fixed values of $H_{s1}/J$ and $T/J$, and increasing the bulk field $H/J$. \\
Hereafter, the surface field effect on the parameters $T_{w}/J$ and $\Omega_{w}/J$ will be discussed for $p=1$. The notation $1^{k}O^{N-k}$ is a situation where k layers are spin-up while N-k layers are spin-down from the top surface $k=1$ to the surface of the bottom $k=N$.
In order to study the behaviour of the system under the effect of the temperature, for fixed surface field $H_{s1}/J$ and transverse field, we plot in Fig. 1 the corresponding phase diagrams: (a) in absence of transverse field $\Omega/J=0.0$ and (b) in the presence of a transverse field $\Omega/J=1.0$.
 One can note that the effect of increasing the transverse field is to decrease $T_{w}/J$. 
When increasing the temperature the wetting transverse field $\Omega_{w}/J$ decreases. This is shown in Fig. 2(a) for a lower temperature $T/J=0.5$, and Fig. 2(b) for higher temperature $T/J=2.0$. We can note the absence, in Figs. 1(b) and 2(b), of the transitions: $O^{N} \leftrightarrow 1^{N}$. Indeed, these transitions subsist for $\Omega/J=0.0$ and lower temperature, Figs. 1(a) and 2(a) respectively, in agreement with the ground state [33]. \\    
Examining the magnetisation behaviour we found that it decreases rapidly in the transition zone either when increasing the temperature as it is shown in Fig. 3(a) for $\Omega/J=1.0$, or when increasing the transverse field as it is plotted in Fig. 3(b) for $T/J=0.02$. Positive magnetisations, in these figures, are calculated for a bulk field \(H/J \rightarrow 0^{+}\), whereas the negative values correspond to \(H/J \rightarrow 0^{-}\). \\
In order to study the effect of a random transverse field $(0<p<1)$ on the wetting and layering transitions, we introduce the probability law given in $Eq. (3)$. The corresponding phase diagrams in the space $(T/J, \Omega /J, p)$ , are plotted in Fig. 4, for $H_{s1}/J=0.95$. It is important to note here the existence of a critical value $pc$, of the probability $p$, above which the wetting transitions disappears. The numerical value corresponding to Fig. 4 is $p_{c}=0.0885$. This critical probability $p_{c}$ depends on the surface field $(H_{s1}/J)$. As it is illustrated in Fig. 5, $p_{c}$ decreases linearly for values of $H_{s1}/J < 0.85$, and almost linearly for $H_{s1}/J > 0.85$, with a discontinuity point for $H_{s1}/J = 0.85$. A similar result was found by Harris [34], for a disordered system with bond and site dilutions, showing that the critical transverse field at zero temperature presents a discontinuity as the concentration passes through the critical percolation concentration. \\
Below, we will consider the effect of the surface field, and the probability $p$, on the wetting temperature $T_{w}/J$ and the wetting transverse field $\Omega_{w}/J$. 
 Indeed, $T_{w}/J$ vanishes as well as increasing the surface field at fixed probability value, Fig. 6(a), or increasing the probability at fixed surface field value, Fig. 6(b).  
While $\Omega_{w}/J$ does not exhibit the same behaviour. However, it decreases when increasing the surface field for a fixed probability value $p$, Fig. 7(a). Whereas it diverges for a fixed surface field value for sufficiently small value of $p$ as it is illustrated in Fig. 7(b). Furthermore, these figures show that the surface field value making $T_{w}/J=0$ increases for decreasing $p$, whereas the surface field value for which $\Omega_{w}/J=0$ is close to 1.
  
\section{Conclusion}

\mbox{  } Within the effective field theory (EFT) and using the differential operator technique, we have studied the phase diagrams of wetting and layering transitions
of a spin$-1/2$ Ising model, for a random transverse field (RTF). It is found that, in the pure case $p=1$, this system exhibits the same behaviour as the mean field study [33]. The dependency of the wetting temperature and wetting transverse field on the surface field and the probability of the presence of a transverse field, was investigated. We have showed the existence of a critical probability $p_{c}$ above which the wetting and layering transitions disappear. 

\newpage \noindent{\bf References}

\begin{enumerate}

\item[{[1]}] M. J. de Oliveira  and R. B. Griffiths , Surf. Sci. {\bf 71}, 687 (1978).
\item[{[2]}] R. Pandit, M. Schick and M. Wortis, Phys. Rev. B {\bf 26}, 8115 (1982).
\item[{[3]}] M. P. Nightingale, W. F. Saam and M. Schick, Phys. Rev. B {\bf 30},3830 (1984).
\item[{[4]}] A. Patrykiejew A., D. P. Landau and K. Binder, Surf. Sci. {\bf 238}, 317 (1990).
\item[{[5]}] C. Ebner, C. Rottman and M. Wortis, Phys. Rev. B {\bf 28},4186  (1983).  
\item[{[6]}] C. Ebner and W. F. Saam, Phys. Rev. Lett. {\bf 58},587 (1987).
\item[{[7]}] C. Ebner and W. F. Saam, Phys. Rev. B {\bf 35},1822 (1987).
\item[{[8]}] C. Ebner, W. F. Saam and A. K. Sen, Phys. Rev. B {\bf 32},1558 (1987).
\item[{[9]}] S. Ramesh and J. D. Maynard, Phys. Rev. Lett. {\bf 49},47 (1982).
\item[{[10]}] S. Ramesh, Q. Zhang, G. Torso and J. D. Maynard, Phys. Rev. Lett. {\bf 52},2375 (1984).
\item[{[11]}] M. Sutton, S. G. J. Mochrie  and R. J. Birgeneou, Phys. Rev. Lett. {\bf 51},407 (1983);\\
S. G. J. Mochrie, M. Sutton, R. J. Birgeneou, D. E. Moncton and P. M. Horn, Phys. Rev. B {\bf 30},263 (1984).
\item[{[12]}] S. K. Stija, L. Passel, J. Eckart, W. Ellenson and H. Patterson, Phys. Rev. Lett. {\bf 51},411 (1983).
\item[{[13]}] C. Ebner and W. F. Saam, Phys. Rev. A {\bf 22},2776 (1980);\\
     ibid, Phys. Rev. A {\bf 23},1925 (1981);\\
     ibid, Phys. Rev. B {\bf 28},2890 (1983).
\item[{[14]}] D. A. Huse , Phys. Rev. B {\bf 30},1371 (1984). 
\item[{[15]}] S. Dietrich {\it Phase transitions and critical phenomena} Vol. 12, C. Domb and Lebowitz Eds., Academic press London and Orlando (1988).  
\item[{[16]}] H. Nakanishi and M. E. Fisher, J. Chem. Phys. {\bf 78},3279  (1983).
\item[{[17]}] P.G. de Gennes, Solid State Commun. {\bf 1}, 132 (1963).
\item[{[18]}]  A. Benyoussef, H. Ez-Zahraouy and M. Saber, Physica A, {\bf 198}, 593
(1993). 
\item[{[19]}]  A. Benyoussef and H. Ez-Zahraouy, Phys. Stat. Sol. (b) {\bf 179}, 521 (1993). 
\item[{[20]}]  A. Bassir, C.E. Bassir, A. Benyoussef, A. Klumper and J. Zittart, Physica A, {\bf 253}, 473 (1998).
\item[{[21]}]  A. Benyoussef, H. Ez-Zahraouy, J. Phys.: Cond. Matter {\bf 6}, 3411 (1994).    
\item[{[22]}]  T. Kaneyoshi, Phys. Rev. {\bf B 33}, 526 (1986).
\item[{[23]}]  F. C. Sa Barreto and I. P. Fittipaldi, Physica A, {\bf 129},
360 (1985).
\item[{[24]}]  T. Kaneyoshi, E. F. Sarmento and I. P. Fittipaldi, Phys. Stat.
Sol. (b) {\bf 150}, 261 (1988).
\item[{[25]}]  T. Kaneyoshi, E. F. Sarmento and I. P. Fittipaldi, Phys. Rev. 
{\bf B 38}, 2649 (1988).
\item[{[26]}]  A. B. Harris, C. Micheletti and J. Yeomans, J. Stat. Phys. 
{\bf 84},323 (1996).
\item[{[27]}]  A. Benyoussef and H. Ez-Zahraouy, Physica A, {\bf 206}, 196
(1994).
\item[{[28]}]  A. Benyoussef and H. Ez-Zahraouy, J. Phys. {\it I } France 
{\bf 4}, 393 (1994).
\item[{[29]}] R. B. Griffiths, Phys. Rev. Lett. {\bf 23}, 17 (1969).
\item[{[30]}] A. B. Harris, Phys. Rev. {\bf B 12}, 203 (1975).
\item[{[31]}] T.F. Cassol, W. Figueiredo and J.A. Plascak, Phys. Lett. {\bf A 160}, 518 (1991).
\item[{[32]}] Y. Q. Wang and Z.Y. Li, J. Phys.: Cond. Matt. {\bf 6}, 10067 (1994).
\item[{[33]}]  L. Bahmad, A. Benyoussef, A. Boubekri, and H. Ez-Zahraouy,
Phys. Stat. Sol. (b) {\bf 215}, 1091 (1999).
\item[{[34]}]  A. B. Harris J. Phys.: C {\bf 7}, 3082 (1974). 
\item[{[35]}]  N. Boccara, Phys. Lett. {\bf 94 A}, 185 (1983). 
\item[{[36]}]  A. Benyoussef and N. Boccara, J. Phys. {\bf C 16}, 1143 (1983). 
\item[{[37]}]  A. Benyoussef and H. Ez-Zahraouy, Physica Scripta {\bf 57}, 603 (1998). 
\item[{[38]}]  A. Benyoussef and H. Ez-Zahraouy, Phys. Rev. {\bf B 52}, 4245 (1995). 
\end{enumerate}

\newpage \noindent{\bf Figure Captions}\newline
\newline

\noindent{\bf Figure 1}: Phase diagram in the $(H/J,T/J)$ plane for a pure
system (p=1.0) with the surface fields $(H_{s1}/J=0.998,H_{s2}/J=-H_{s1}/J)$ 
a) $\Omega/J=0.0$,
b) $\Omega/J=1.0$.

\noindent{\bf Figure 2}: Phase diagram in the $(H/J, \Omega /J)$ plane for
a pure system (p=1.0)  with $(H_{s1}/J=0.998,H_{s2}/J=-H_{s1}/J)$ 
a) $T/J=0.5$,
b) $T/J=2.0$.

\noindent{\bf Figure 3}: Magnetisations for $(H_{s1}/J=0.95,H_{s2}/J=-H_{s1}/J)$: \\
a) as function of the temperature for $\Omega /J=1.0$, \\
b) as function of the transverse field for $T /J=0.02$. 

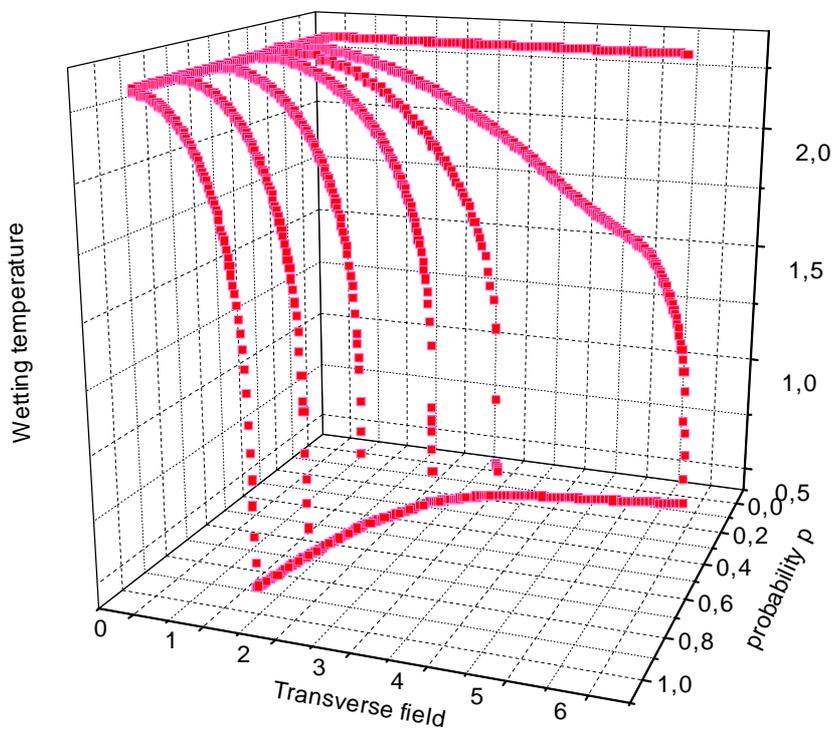
\noindent{\bf Figure 4}: Phase diagrams in the $(T/J, \Omega/J ,p)$
space for $(H_{s1}/J=0.95,H_{s2}/J=-H_{s1}/J)$.

\noindent {\bf Figure 5}: The critical probability $p_{c}$ dependence on
the surface field $H_{s1}/J$ for $T/J=0.02$.

\noindent {\bf Figure 6}: Wetting temperature $T_{w}/J$ as a function of: \\
a) surface field $H_{s1}/J$ for $\Omega/J=1.0$ and several values of the probability p \\
b) probability p for $\Omega/J=2.0$ and several values of the surface field $H_{s1}/J$. 

\noindent {\bf Figure 7}: Wetting transverse field $\Omega _{w}/J$ as a
function of : \\
a) surface field $H_{s1}/J$ for $T/J=0.02$ and several values of the probability p \\
b) probability p for $T/J=2.0$ and several values of the surface field $H_{s1}/J$.

\newpage \appendix
%\chapter{ \bf Appendix A}

\section{Expression of the coefficients $A_{k}(\mu)$}

%,(\mu =\Omega ,0), k=0,...,19$}
Using the functions $f_\mu(x), (\mu =\Omega ,0)$ and the field $H_{i}$,
defined in the body text, the coefficients $A_{k}(\mu)(\mu =\Omega ,0),
k=0,...,19$ for a layer i, are as follows :

\begin{equation}
\begin{array}{l}
A_{0}(\mu)=(1/2)^6\times(f_\mu(6J+H_{i})+6f_\mu(4J+H_{i})+15f_\mu(2J+H_{i})+
\\ 
20f_\mu(H_{i})+15f_\mu(-2J+H_{i})+6f_\mu(-4J+H_{i})+f_\mu(-6J+H_{i}))
\end{array}
\end{equation}

\begin{equation}
\begin{array}{l}
A_{1}(\mu)=4\times(1/2)^6\times(f_\mu(6J+H_{i})+4f_\mu(4J+H_{i})+5f_%
\mu(2J+H_{i})- \\ 
5f_\mu(-2J+H_{i})-4f_\mu(-4J+H_{i})-f_\mu(-6J+H_{i}))
\end{array}
\end{equation}

\begin{equation}
\begin{array}{l}
A_{2}(\mu)=6\times(1/2)^6\times(f_\mu(6J+H_{i})+2f_\mu(4J+H_{i})-2f_%
\mu(2J+H_{i})- \\ 
4f_\mu(H_{i})-f_\mu(-2J+H_{i})+2f_\mu(-4J+H_{i})+f_\mu(-6J+H_{i}))
\end{array}
\end{equation}

\begin{equation}
\begin{array}{l}
A_{3}(\mu)=4\times(1/2)^6\times(f_\mu(6J+H_{i})-3f_\mu(2J+H_{i})+3f_%
\mu(-2J+H_{i})- \\ 
f_\mu(-6J+H_{i}))
\end{array}
\end{equation}

\begin{equation}
\begin{array}{l}
A_{4}(\mu)=(1/2)^6\times(f_\mu(6J+H_{i})-2f_\mu(4J+H_{i})-f_\mu(2J+H_{i})+
\\ 
4f_\mu(H_{i})-f_\mu(-2J+H_{i})-2f_\mu(-4J+H_{i})+f_\mu(-6J+H_{i}))
\end{array}
\end{equation}

\begin{equation}
\begin{array}{l}
A_{5}(\mu)=(1/2)^6\times(f_\mu(6J+H_{i})+4f_\mu(4J+H_{i})+5f_\mu(2J+H_{i})-
\\ 
5f_\mu(-2J+H_{i})-4f_\mu(-4J+H_{i})-f_\mu(-6J+H_{i}))
\end{array}
\end{equation}

\begin{equation}
\begin{array}{l}
A_{6}(\mu)=4\times(1/2)^6\times(f_\mu(6J+H_{i})+2f_\mu(4J+H_{i})-f_%
\mu(2J+H_{i})- \\ 
4f_\mu(H_{i})-5f_\mu(-2J+H_{i})+2f_\mu(-4J+H_{i})+f_\mu(-6J+H_{i}))
\end{array}
\end{equation}

\begin{equation}
\begin{array}{l}
A_{7}(\mu)=6\times(1/2)^6\times(f_\mu(6J+H_{i})-3f_\mu(2J+H_{i})+3f_%
\mu(-2J+H_{i})- \\ 
f_\mu(-6J+H_{i}))
\end{array}
\end{equation}

\begin{equation}
\begin{array}{l}
A_{8}(\mu)=4\times(1/2)^6\times(f_\mu(6J+H_{i})-2f_\mu(4J+H_{i})-f_%
\mu(2J+H_{i})- \\ 
4f_\mu(H_{i})-f_\mu(-2J+H_{i})-2f_\mu(-4J+H_{i})+f_\mu(-6J+H_{i}))
\end{array}
\end{equation}

\begin{equation}
\begin{array}{l}
A_{9}(\mu)=(1/2)^6\times(f_\mu(6J+H_{i})-4f_\mu(4J+H_{i})+5f_\mu(2J+H_{i})-
\\ 
5f_\mu(-2J+H_{i})+4f_\mu(-4J+H_{i})-f_\mu(-6J+H_{i}))
\end{array}
\end{equation}

\begin{equation}
\begin{array}{l}
A_{10}(\mu)=(1/2)^6\times(f_\mu(6J+H_{i})+4f_\mu(4J+H_{i})+5f_\mu(2J+H_{i})-
\\ 
5f_\mu(-2J+H_{i})-4f_\mu(-4J+H_{i})-f_\mu(-6J+H_{i}))
\end{array}
\end{equation}

\begin{equation}
\begin{array}{l}
A_{11}(\mu)=4\times(1/2)^6\times(f_\mu(6J+H_{i})+2f_\mu(4J+H_{i})-f_%
\mu(2J+H_{i})- \\ 
4f_\mu(H_{i})-f_\mu(-2J+H_{i})+2f_\mu(-4J+H_{i})+f_\mu(-6J+H_{i}))
\end{array}
\end{equation}

\begin{equation}
\begin{array}{l}
A_{12}(\mu)=6\times(1/2)^6\times(f_\mu(6J+H_{i})-3f_\mu(2J+H_{i})+3f_%
\mu(-2J+H_{i})- \\ 
f_\mu(-6J+H_{i}))
\end{array}
\end{equation}

\begin{equation}
\begin{array}{l}
A_{13}(\mu)=4\times(1/2)^6\times(f_\mu(6J+H_{i})-2f_\mu(4J+H_{i})-f_%
\mu(2J+H_{i})+ \\ 
4f_\mu(H_{i})-f_\mu(-2J+H_{i})-2f_\mu(-4J+H_{i})+f_\mu(-6J+H_{i}))
\end{array}
\end{equation}

\begin{equation}
\begin{array}{l}
A_{14}(\mu)=(1/2)^6\times(f_\mu(6J+H_{i})-4f_\mu(4J+H_{i})+5f_\mu(2J+H_{i})-
\\ 
5f_\mu(-2J+H_{i})+4f_\mu(-4J+H_{i})-f_\mu(-6J+H_{i}))
\end{array}
\end{equation}

\begin{equation}
\begin{array}{l}
A_{15}(\mu)=(1/2)^6\times(f_\mu(6J+H_{i})+2f_\mu(4J+H_{i})-f_\mu(2J+H_{i})-
\\ 
4f_\mu(H_{i})-f_\mu(-2J+H_{i})+2f_\mu(-4J+H_{i})+f_\mu(-6J+H_{i}))
\end{array}
\end{equation}

\begin{equation}
\begin{array}{l}
A_{16}(\mu)=4\times(1/2)^6\times(f_\mu(6J+H_{i})-3f_\mu(2J+H_{i})+3f_%
\mu(-2J+H_{i})- \\ 
f_\mu(-6J+H_{i}))
\end{array}
\end{equation}

\begin{equation}
\begin{array}{l}
A_{17}(\mu)=6\times(1/2)^6\times(f_\mu(6J+H_{i})-2f_\mu(4J+H_{i})-f_%
\mu(2J+H_{i})+ \\ 
4f_\mu(H_{i})-f_\mu(-2J+H_{i})-2f_\mu(-4J+H_{i})+f_\mu(-6J+H_{i}))
\end{array}
\end{equation}

\begin{equation}
\begin{array}{l}
A_{18}(\mu)=4\times(1/2)^6\times(f_\mu(6J+H_{i})-4f_\mu(4J+H_{i})+5f_%
\mu(2J+H_{i})- \\ 
5f_\mu(-2J+H_{i})+4f_\mu(-4J+H_{i})-f_\mu(-6J+H_{i}))
\end{array}
\end{equation}

\begin{equation}
\begin{array}{l}
A_{19}(\mu)=(1/2)^6\times(f_\mu(6J+H_{i})-6f_\mu(4J+H_{i})+15f_\mu(2J+H_{i})-
\\ 
20f_\mu(H_{i})+15f_\mu(-2J+H_{i})-6f_\mu(-4J+H_{i})+f_\mu(-6J+H_{i}))
\end{array}
\end{equation}

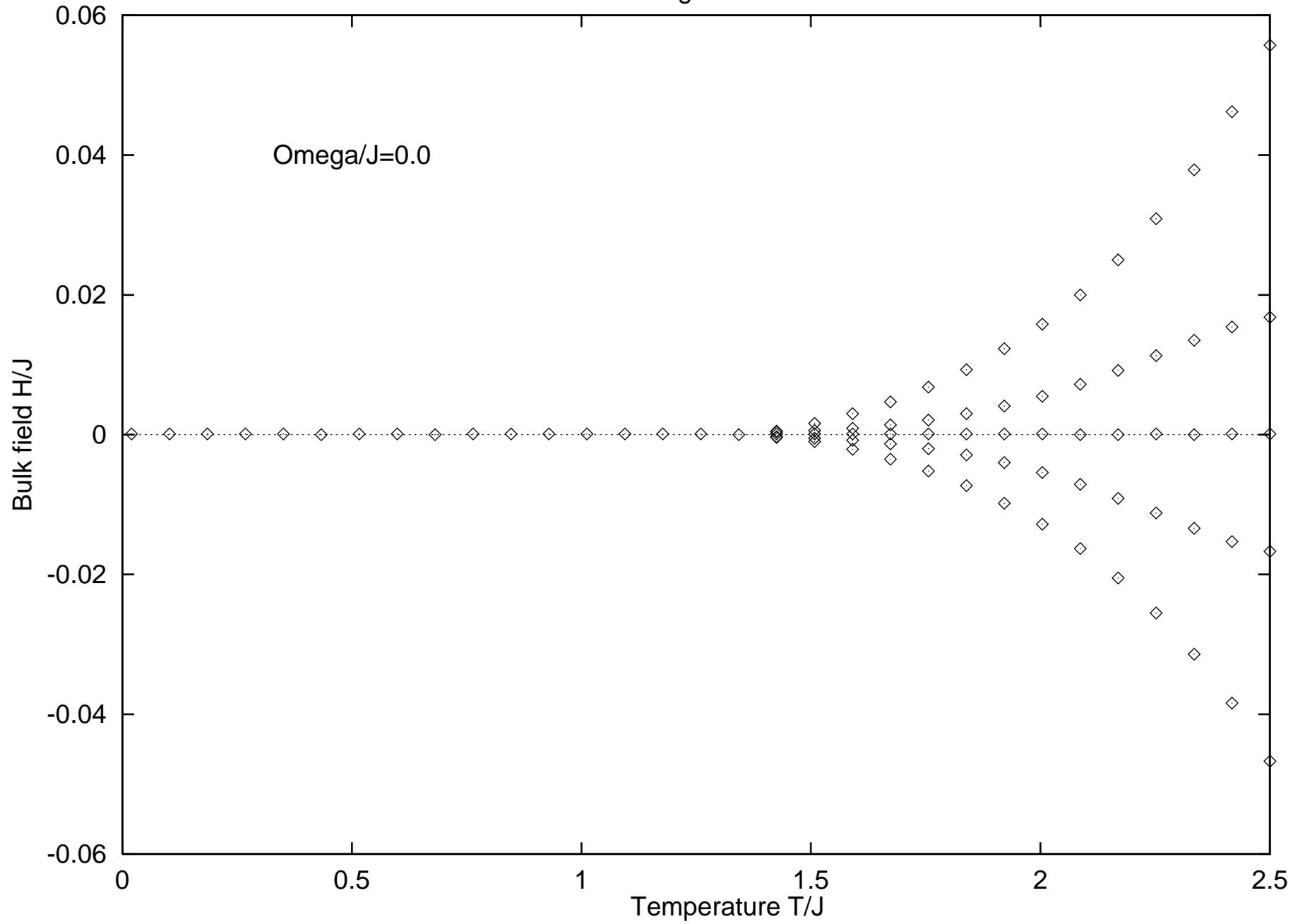
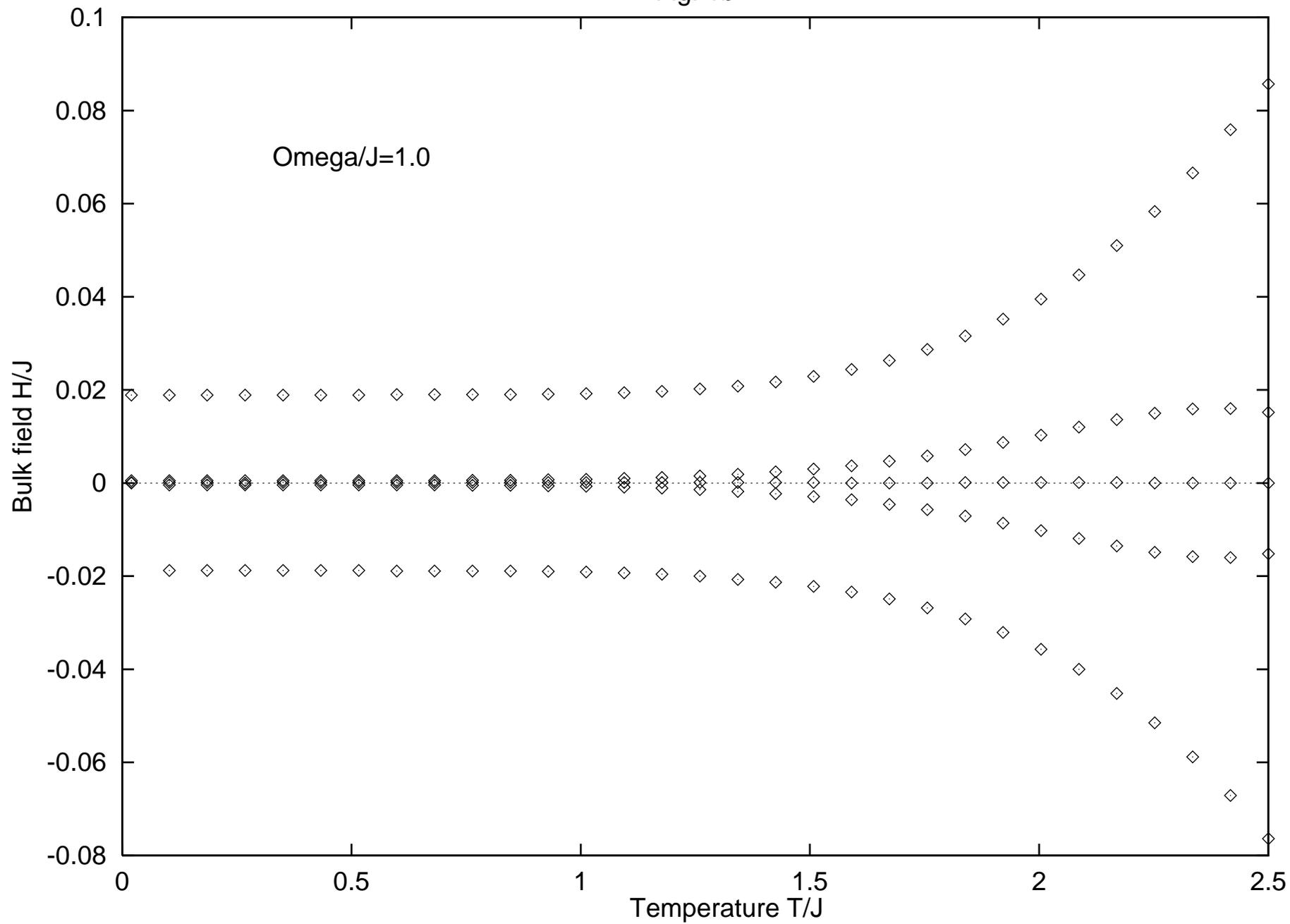
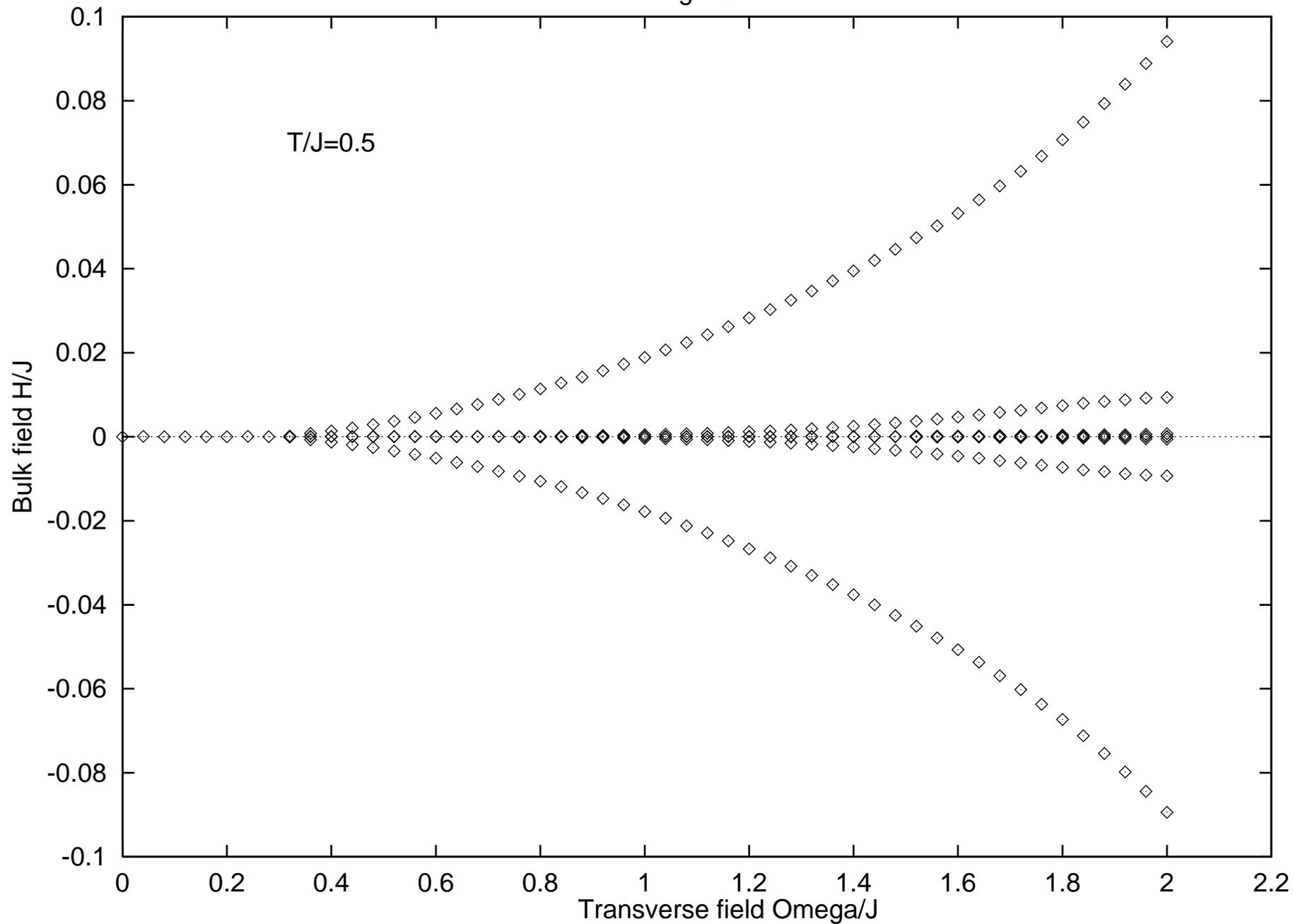
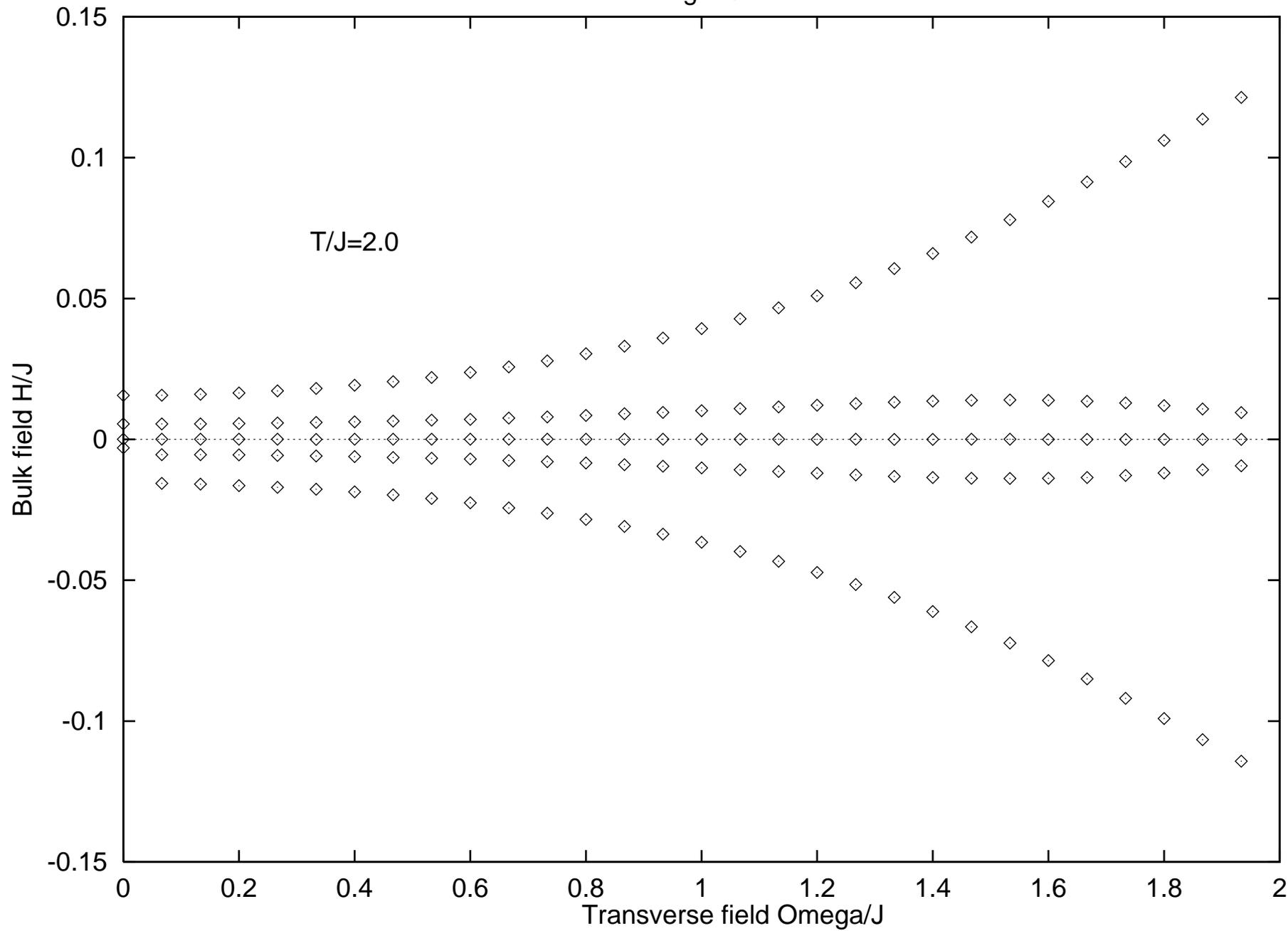
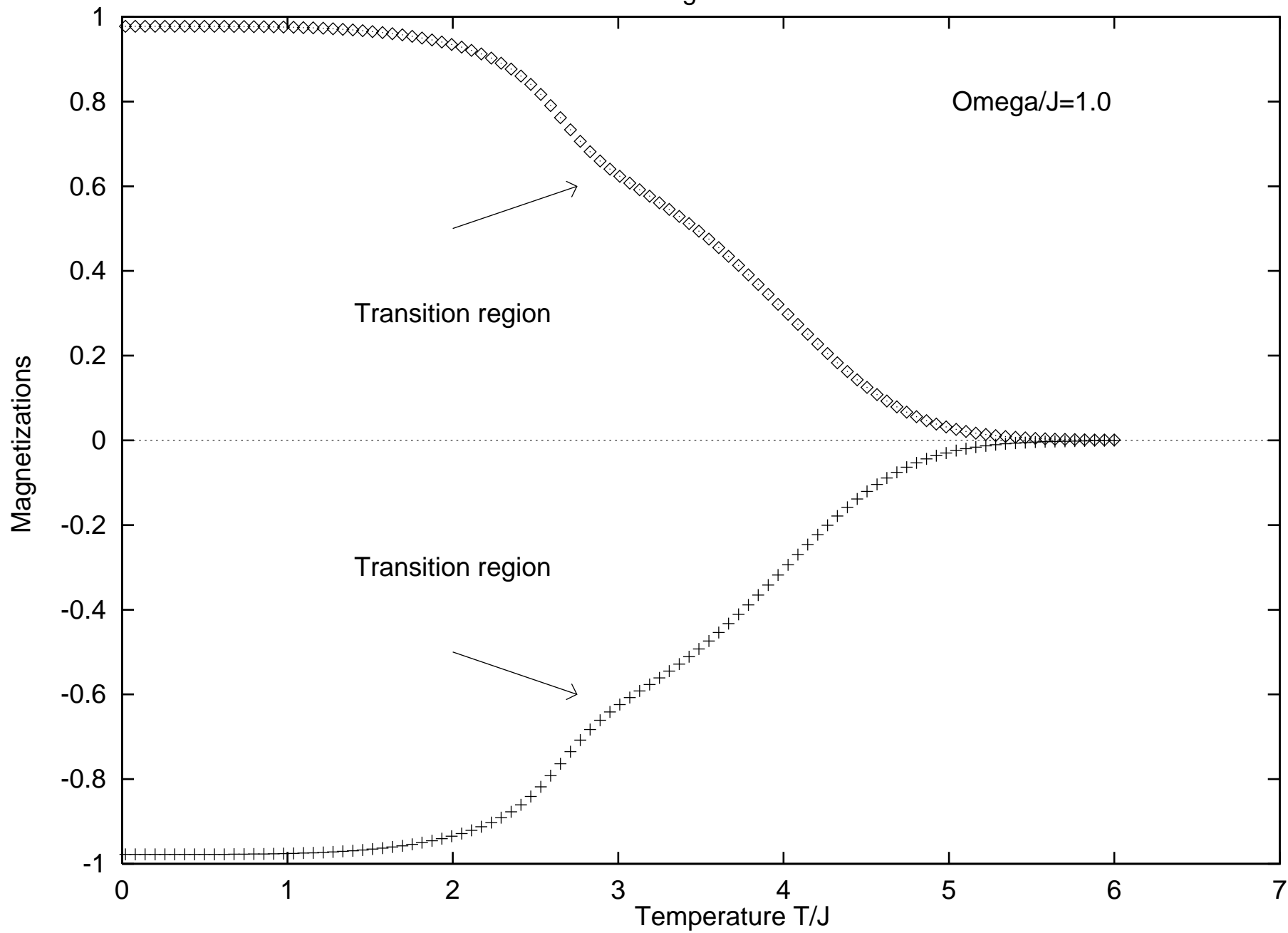
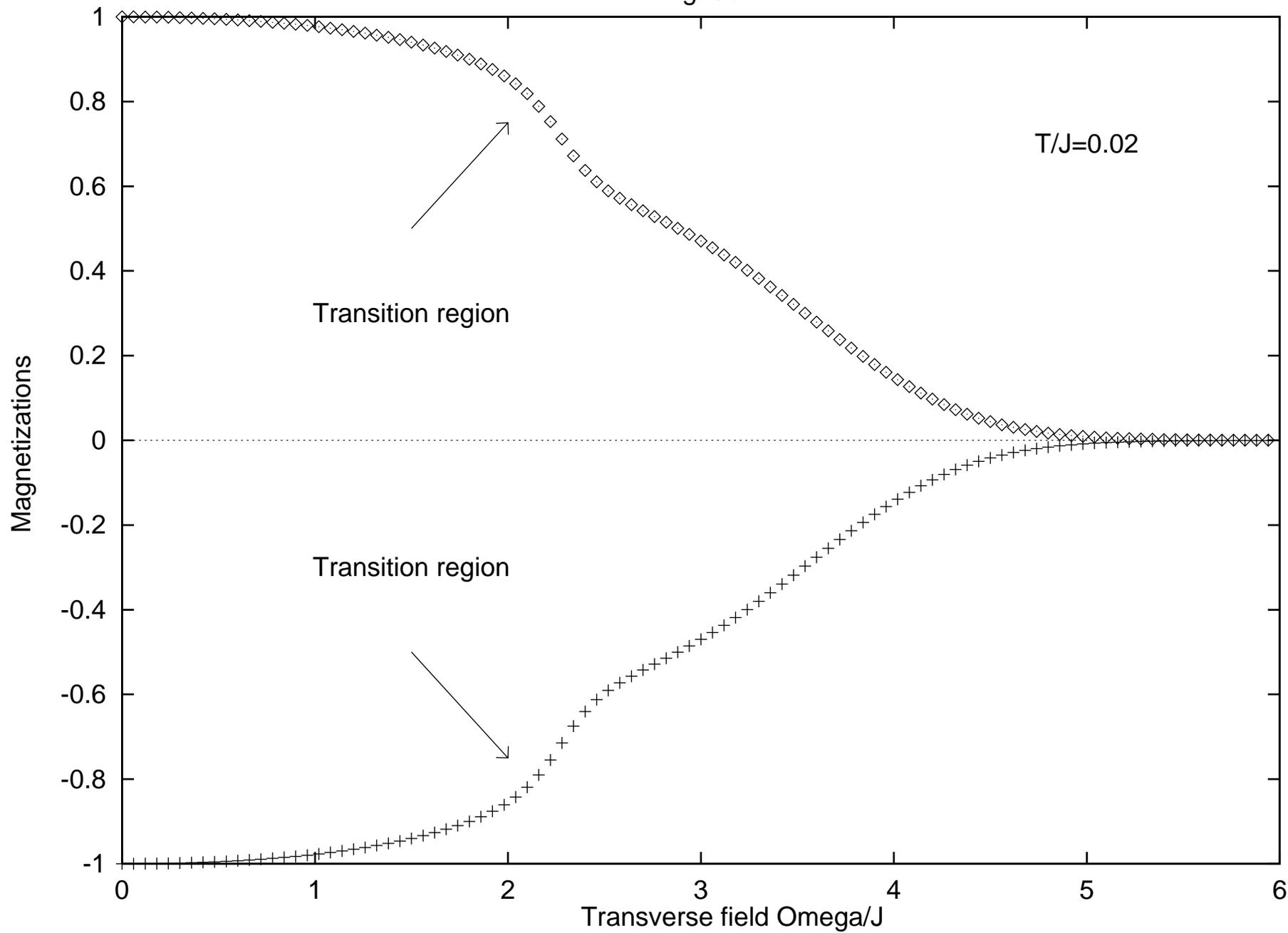
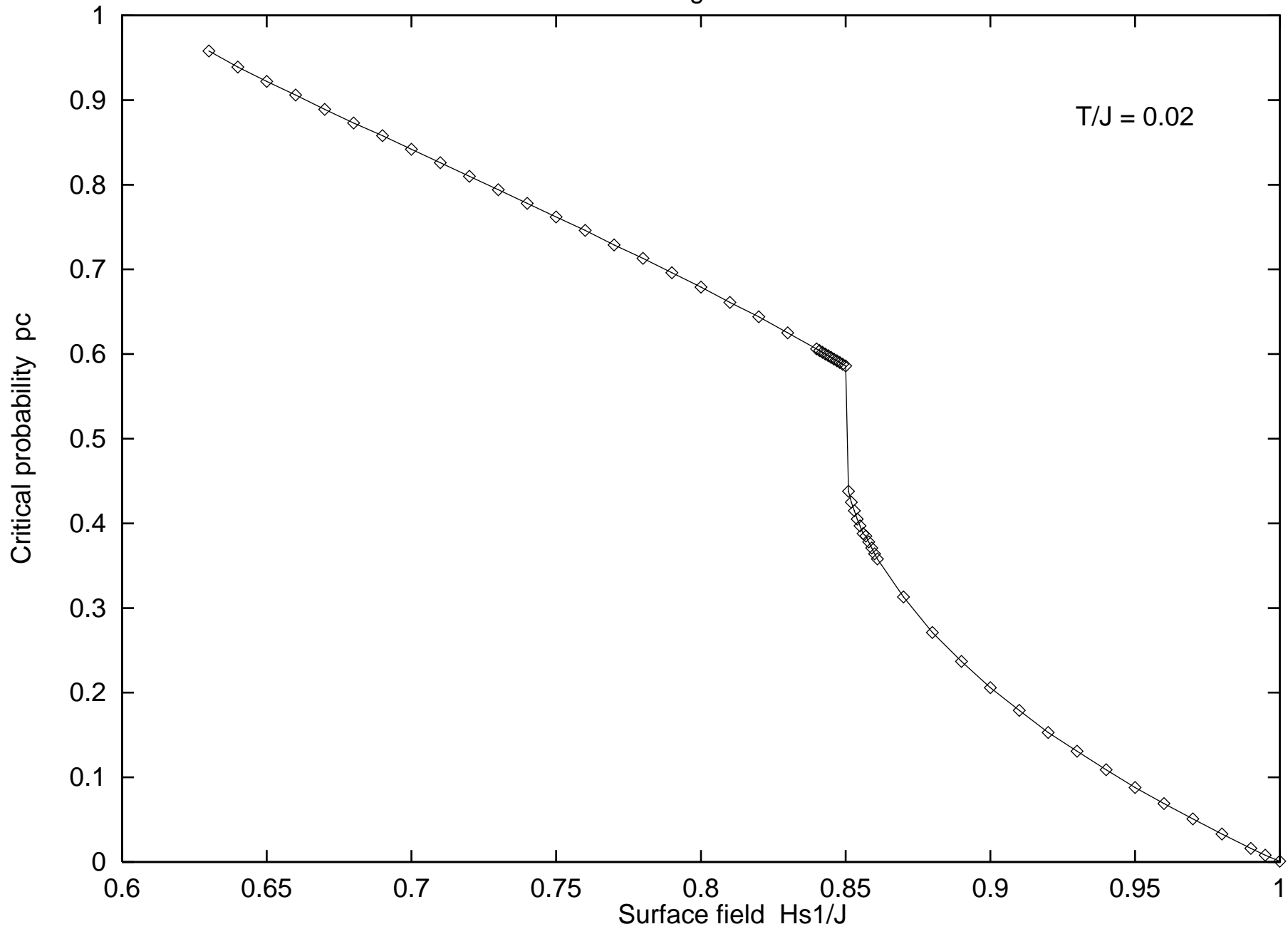
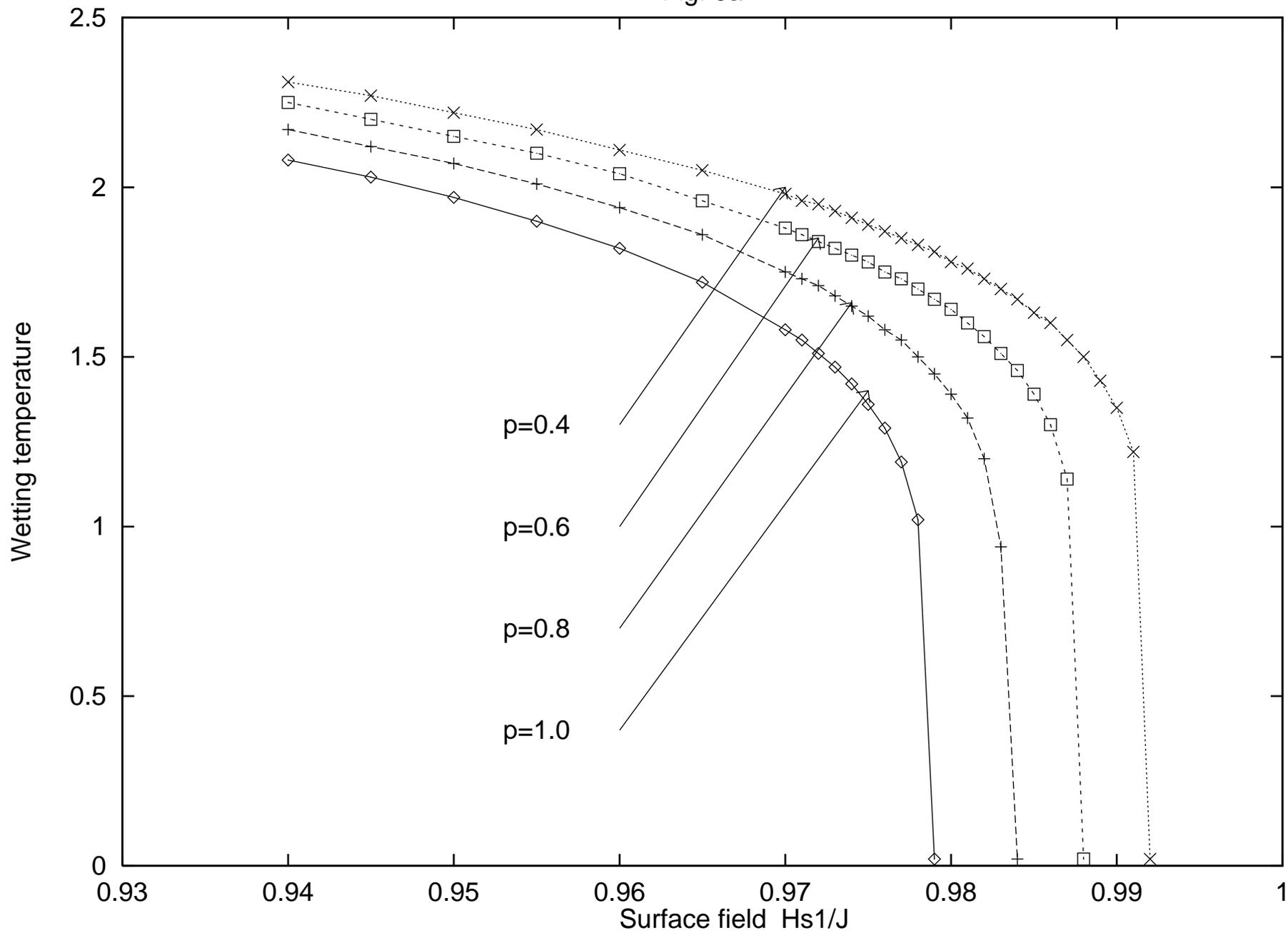
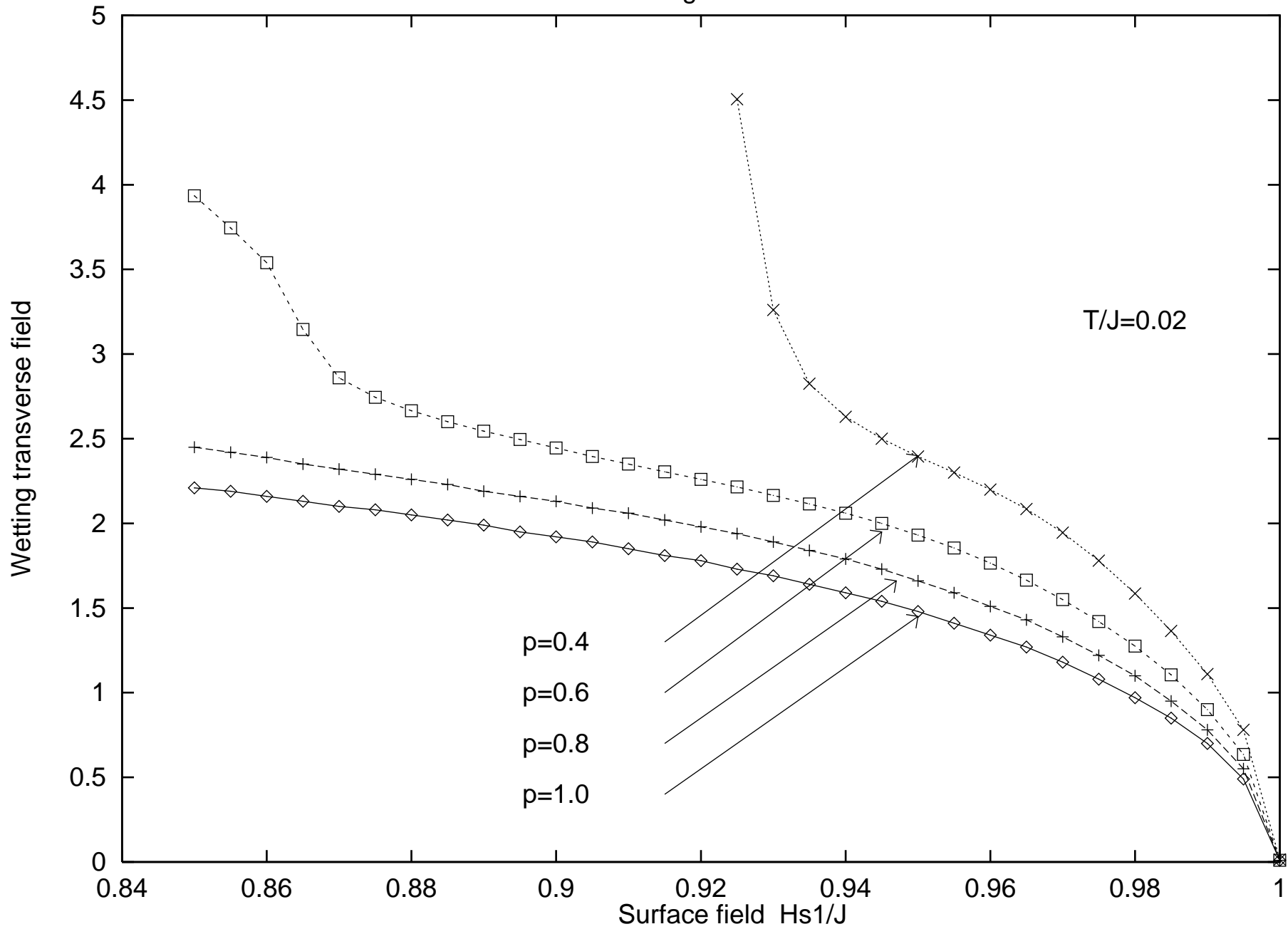
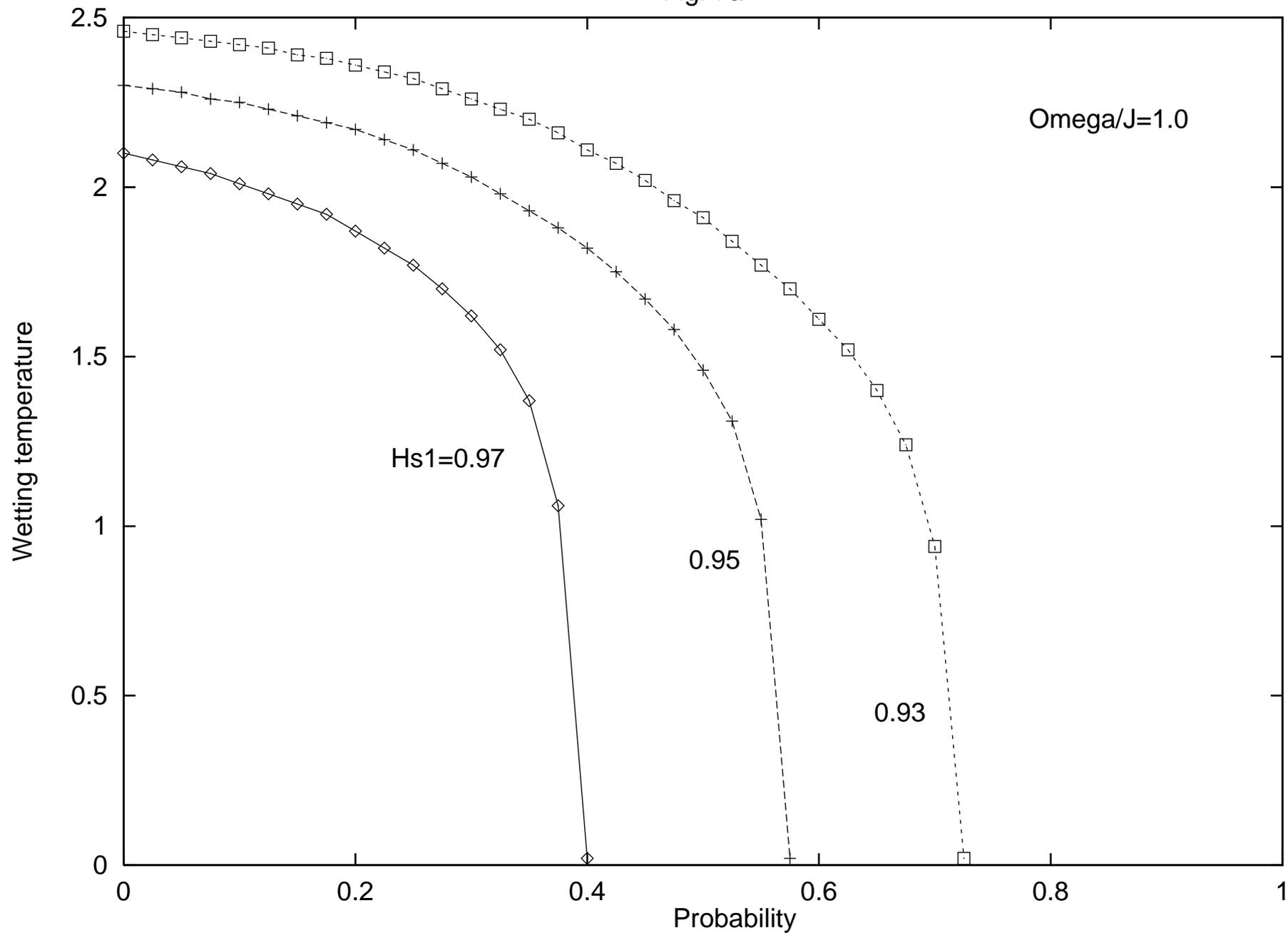
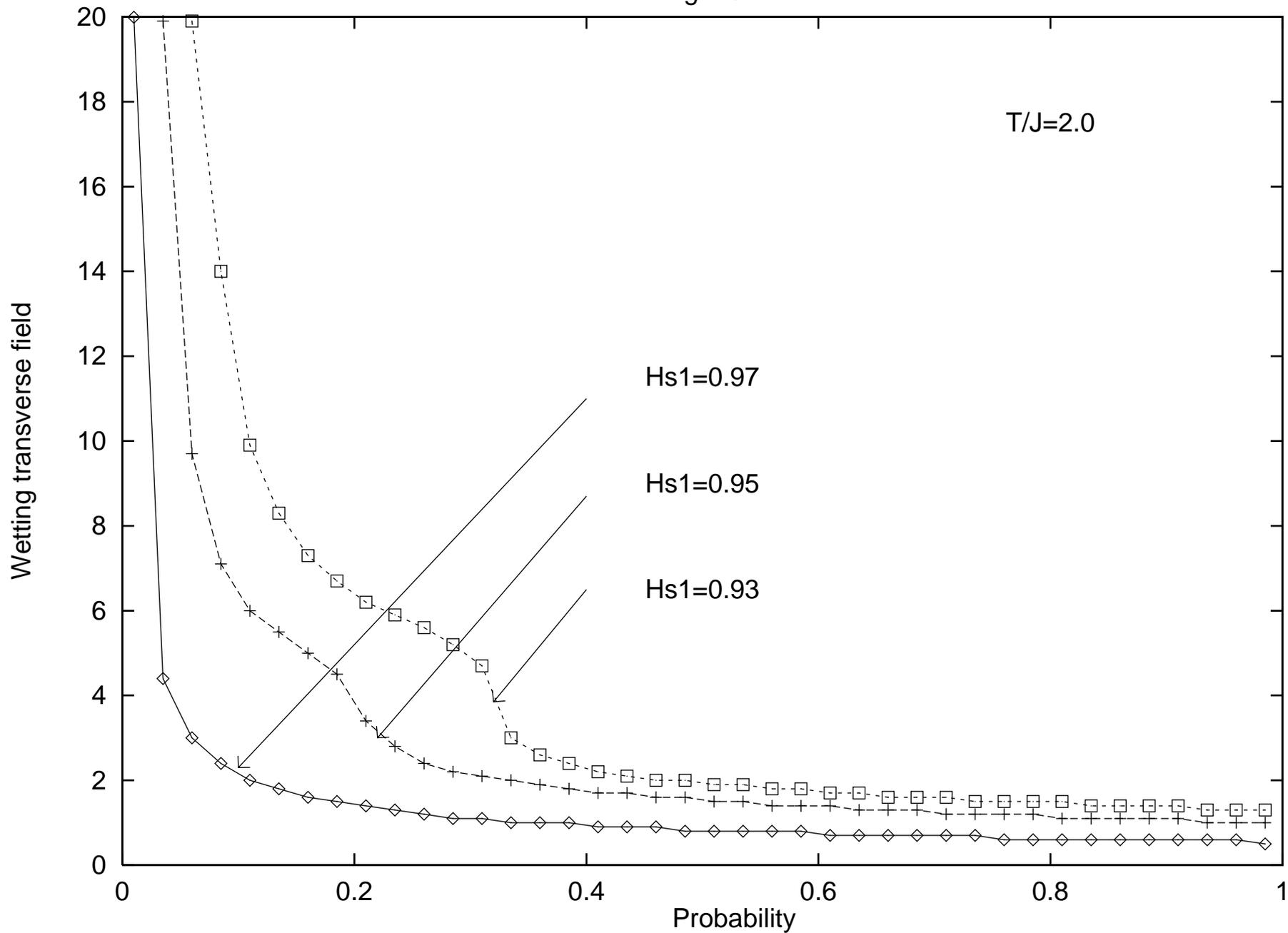
\end{document}